\documentclass[12pt]{iopart}

\usepackage{graphics}
\usepackage{amssymb}
\usepackage{bm}

\begin{document}

\title{Nonlinear electrodynamics and CMB polarization}

\author{Herman J. Mosquera Cuesta$^{1,2,3}$ and G. Lambiase$^{4,5}$}

\address{\mbox{$^1$Departmento de F{\'\i}sica Universidade Estadual Vale do Acara\'u} \\
\mbox{Avenida da Universidade 850, Campus da Bet\^ania, CEP 62.040-370, Sobral, Cear\'a,
Brazil} \\
{$^2$Instituto de Cosmologia, Relatividade e Astrof\'\i sica  (ICRA-BR) \\
Centro Brasileiro de Pesquisas F{\'\i}sicas} \\
\mbox{Rua Dr. Xavier Sigaud 150, CEP 22290-180, Urca Rio de Janeiro, RJ, Brazil} \\
\mbox{$^3$International Center for Relativistic Astrophysics Network (ICRANet)} \\
\mbox{International Coordinating Center, Piazzalle della Repubblica 10, 065112, Pescara, Italy}\\
$^4$Dipartimento di Fisica "E. R.Caianiello", Universit\'a di Salerno, 84081 Baronissi (Sa), Italy \\
$^5$INFN, Sezione di Napoli, Italy
}

\ead{herman@icra.it, lambiase@sa.infn.it}

\begin{abstract}

Recently WMAP and BOOMERanG experiments have set stringent constraints on the
polarization angle of photons propagating in an expanding universe: $\Delta \alpha = (-2.4
\pm 1.9)^\circ$. The polarization of the Cosmic Microwave Background radiation (CMB)
is reviewed in the context of nonlinear electrodynamics (NLED). We compute the polarization
angle of photons propagating in a cosmological background with planar symmetry. For this
purpose, we use the Pagels-Tomboulis (PT) Lagrangian density describing NLED, which has
the form $L\sim  (X/\Lambda^4)^{\delta - 1}\; X $, where $X=\frac{1}{4}F_{\alpha\beta}
F^{\alpha \beta}$, and $\delta$
the parameter featuring the non-Maxwellian character of the PT nonlinear description
of the electromagnetic interaction. After looking at the polarization components in
the plane orthogonal to the ($x$)-direction of propagation of the CMB photons, the
polarization angle is defined in terms of the eccentricity of the universe, a
geometrical property whose evolution on cosmic time (from the last scattering
surface to the present) is constrained by the strength of magnetic fields over
extragalactic distances.
\end{abstract}

\maketitle

\section{Introduction}


Modifications to the standard (Maxwell) electrodynamics were proposed
in the literature in order to avoid infinite physical quantities from
theoretical descriptions of electromagnetic interactions. Born and Infeld
\cite{born}, for instance, proposed a model in which the infinite self
energy of point particles (typical of Maxwell's electrodynamics) are
removed by introducing an upper limit on the electric field strength,
and by considering the electron as an
electric particle with finite radius. Along this line, other models of
nonlinear electrodynamics (NLED) Lagrangians were proposed by Plebanski,
who also showed that Born-Infeld model satisfies physically acceptable
requirements \cite{plebanski}. Consequences of nonlinear
electrodynamics have been studied in many contexts, such a, for
example, cosmological models \cite{cosmology}, black holes and
wormhole physics \cite{BH,wormhole}, primordial magnetic
fields in the Universe \cite{kunze,cea,cuesta-lambiase},
gravitational baryogenesis \cite{cuesta-lambiase}, and
astrophysics \cite{applications,cuesta}.

In this paper we investigate the CMB polarization of photons described
by nonlinear electrodynamics. We compute the polarization angle of
photons propagating in an expanding Universe, by considering in particular
cosmological models with planar symmetry. The polarization angle does depend
on the parameter characterizing the nonlinearity of electrodynamics, which
will be constrained by making use of the recent data from WMAP and BOOMERANG.
This kind of investigations has received a lot of interest because they
represent a probe of models beyond the standard model, which may violate the
fundamental symmetries such as CPT and Lorentz invariance \cite{zhang,amelinojcap}.
In what follows we will follow the main lines of the paper on ``Cosmological
CPT violation, baryo/leptogenesis and CMB polarization'' by Li-Xia-Li-Zhang \cite{zhang2007}.

\section{Minimally coupling gravity to nonlinear electrodynamics}

The action of (nonlinear) electrodynamics coupled minimally to gravity is

 \begin{equation}\label{action}
S=\frac{1}{2\kappa}\int d^4x \sqrt{-g}R+\frac{1}{4\pi}\int d^4x \sqrt{-g}L(X, Y)\,,
 \end{equation}

where $\kappa=8\pi G$, $L$ is the Lagrangian of nonlinear electrodynamics
depending on the invariant $X=\frac{1}{4}F_{\mu\nu}F^{\mu\nu} = -2({\bf E}^2-{\bf B}^2)$
and $Y= \frac{1}{4}F_{\mu\nu}\, {^*}F^{\mu\nu}$, where $F^{\mu\nu} \equiv \nabla_\mu
A_\nu - \nabla_\nu A_\mu$, and ${^*}F^{\mu\nu}=\epsilon^{\mu\nu\rho\sigma}
F_{\rho\sigma}$ is the dual bivector, and $\epsilon^{\alpha\beta\gamma\delta}
= \frac{1}{2\sqrt{-g}}\; \varepsilon^{\alpha\beta \gamma\delta}$, with
$\varepsilon^{\alpha\beta\gamma\delta}$ the Levi-Civita tensor ($\varepsilon_{
0123} = +1$).

The equations of motion are \cite{kunze}

\begin{equation}\label{eqsofmotion}
    \nabla_\mu\left(-L_X F^{\mu\nu}-L_Y {^*}F^{\mu\nu}\right)=0\,,
\end{equation}

where $L_X=\partial L/\partial X$ and $L_Y=\partial L/\partial Y$,
\begin{equation}\label{bianchi}
  \nabla_\mu F_{\nu\lambda}+\nabla_\nu F_{\lambda\mu}+\nabla_\lambda
  F_{\mu\nu}=0\,.
\end{equation}

After a swift grasp on this set of equations one realizes that is difficult
to find solutions in closed form of these equations.
Therefore to study the effects of nonlinear electrodynamics, we confine
ourselves to consider the abelian Pagels-Tomboulis theory \cite{pagel},
proposed as an effective model of low energy  QCD. The Lagrangian density
of this theory involves only the invariant $X$ in the form

\begin{equation}\label{L(X)}
L(X)=-\left(\frac{X^2}{\Lambda^8}\right)^{\frac{\delta-1}{2}}X
=-\gamma X^\delta\,,
\end{equation}

where $\gamma$ (or $\Lambda$) and $\delta$ are free parameters that,
with appropriate choice, reproduce the well known Lagrangian already
studied in the literature. $\gamma$ has dimensions [energy]$^{4(1-\delta)}$.

Following Kunze \cite{kunze}, the energy momentum tensor corresponding
to the Lagrangian density $L(X)$ is given by

\begin{equation}\label{energy-momentum-tensor}
    T_{\mu\nu}=\frac{1}{4\pi}\left[L_Xg^{\alpha\beta}F_{\mu\alpha}
F_{\beta\nu}+g_{\mu\nu}L\right]
\end{equation}
and the decomposition of the electromagnetic tensor with respect to a
fundamental observer with 4-velocity $u_\mu$ ($u_\mu u^\mu=-1$)
\begin{equation}\label{decompositionF}
F_{\mu\nu}=2{\hat E}_{[\mu} u_{\nu]}-\eta_{\mu\nu\varsigma\tau}
u^\varsigma {\hat B}^\tau\,.
\end{equation}

The electric and magnetic fields are therefore given by ${\hat E}_\mu=
F_{\mu\nu}u^\nu$ and ${\hat B}_\mu=\frac{1}{2}
\eta_{\mu\nu\kappa\lambda}u^\nu F^{\kappa\lambda}$ ($\eta_{\alpha\beta
\gamma\delta}= \sqrt{-g}\; \varepsilon_{\alpha\beta\gamma\delta}$).

The energy density turns out to be

 \begin{equation}\label{energydensity1}
 \rho=T_{\mu\nu}u^\mu u^\nu=-\frac{1}{8\pi}\frac{L}{X}\left[(2\delta-1)
{\hat E}_\alpha {\hat E}^\alpha + {\hat B}_\alpha{\hat B}^\alpha\right]\,.
 \end{equation}

The positivity of $\rho$ (weak energy condition) imposes, in general, the
constraint on $\delta$. For the Lagrangian (\ref{L(X)}) one gets $\delta
\geq \frac{1}{2}$. However, this condition can be relaxed because we shall
consider cosmological scenarios in which the electric field is zero, and
only the magnetic fields survive (this is justified by the fact that during
the radiation dominated era the plasma effects induce a rapid decay of the
electric field, whereas magnetic field remains (see the paper by Turner and
Widrow in \cite{primordialoriginB})).

The equation of motion for the Pagels-Tomboulis theory follows from Eq.
(\ref{eqsofmotion}) with $Y=0$
\begin{equation}\label{eq-motionPagels}
\nabla_\mu F^{\mu\nu}=-(\delta-1)\frac{\nabla_\mu X}{X}\, F^{\mu\nu}\,.
\end{equation}
In terms of the potential vector $A^\mu$, and imposing the Lorentz gauge
$\nabla_\mu A^\mu=0$, Eq. (\ref{eq-motionPagels}) becomes
\begin{equation}\label{eq-A}
\nabla_\mu \nabla^\mu A^\nu+R^\nu_{\,\,\mu}A^\mu=-(\delta-1)\frac{\nabla_\mu
X}{X}\, (\nabla^\mu A^\nu-\nabla^\nu A^\mu)\,,
\end{equation}
where the Ricci tensor $R^\nu_{\,\, \mu}$ appears because the relation
$[\nabla^\mu, \nabla_\nu]A^\nu=-R^\mu_{\,\, \mu}A^\mu$.

To proceed onward, we apply the geometrical optics approximation. This means
that the scales of variation of the electromagnetic fields are smaller than
the cosmological scales we consider next. In this approximation, the 4-vector
$A^\mu(x)$ can be written as \cite{misner}

\begin{equation}\label{A-expansion}
A^\mu(x)={\it Re}\left[(a^\mu(x)+\epsilon b^\mu(x)+\ldots)e^{iS(x)/\epsilon}\right]
\end{equation}
with $\epsilon \ll 1$ so that the phase $S/\epsilon$ varies faster than the
amplitude. By defining the wave vector $k_\mu=\nabla_\mu S$, which defines
the direction of the photon propagation, one finds that the gauge condition
implies $k_\mu a^\mu=0$ and $k_\mu b^\mu=0$. It turns out to be convenient
to introduce the normalized polarization vector $\varepsilon^\mu$ so that
the vector $a^\mu$ can be written as

\begin{equation}\label{a-mu}
a^\mu(x)= A(x) \varepsilon^\mu\,, \qquad \varepsilon_\mu \varepsilon^\mu =1\,.
\end{equation}
As a consequence of (\ref{a-mu}), one also finds $k_\mu \varepsilon^\mu =0$,
i.e. the wave vector is orthogonal to the polarization vector.

By making use of Eq. (\ref{A-expansion}), one obtains
\begin{equation}\label{nablaX/X}
\frac{\nabla_\mu X}{X}=\frac{2ik_\mu}{\epsilon}[1+\Omega(\epsilon)]\,,
\end{equation}
where
 \[
 \Omega\equiv \frac{\epsilon i k_{[\alpha}a_{\beta]}\nabla_\mu
k^{[\alpha}a^{\beta]}+{\cal O}(\epsilon^2)}{-(k_{[\alpha}a_{\beta]})^2
+i\epsilon 2 k_{[\alpha}a_{\beta]}(\nabla^{[\alpha}a^{\beta]}+
ik^{[\alpha}a^{\beta]})+{\cal O}(\epsilon^2)}\,.
 \]

To leading order in $\epsilon$, the term depending on the Ricci tensors
can be neglected in (\ref{eq-A}). Inserting (\ref{A-expansion}) into
(\ref{eq-A}) and collecting all terms proportional to $\epsilon^{-2}$
and $\epsilon^{-1}$, one obtains

\begin{eqnarray}
  \frac{1}{\epsilon^2}: & & k_\mu k^\mu a^\sigma = 2(\delta-1) k_\mu
k^{[\mu}a^{\sigma]}\,, \label{k2=0a} \\
\frac{1}{\epsilon}: & & 2k_\mu \nabla^\mu a^\sigma+a^\sigma\nabla_\mu
k^\mu = -2(\delta-1) k_\mu \nabla^{[\mu} a^{\sigma]}\,.
  \label{k2-delta}
\end{eqnarray}

Taking into account the gauge condition $k^\mu a_\mu=0$, the first equation implies

 \begin{equation}\label{k2=0}
 (2\delta-1)k^2=0 \quad \to \quad  k_\mu k^\mu =0\,, \quad \mbox{provided} \quad\delta
\neq \frac{1}{2}\,,
 \end{equation}

hence photons propagate along null geodesics. Multiplying Eq. (\ref{k2-delta}) by $a_\sigma$,
and using (\ref{a-mu}) one obtains

\[
 \frac{1}{2}\nabla_\mu k^\mu = -\delta k^\mu \nabla_\mu \ln A + (\delta-1)
k_\mu \varepsilon^\sigma \nabla_\sigma \varepsilon^\mu\,,
 \]

so that Eq. (\ref{k2-delta}) can be recast in the form

\begin{equation}\label{k-nabla-epsilon}
k^\mu \nabla_\mu \varepsilon^\sigma = \frac{\delta-1}{\delta}\, \Upsilon^\sigma
\end{equation}

where

 \begin{equation}\label{Asigma}
 \Upsilon^\sigma \equiv k_\mu \left[\nabla^\sigma \varepsilon^\mu-
(\varepsilon^\rho \nabla_\rho \varepsilon^\mu)\varepsilon^\sigma\right]\,.
 \end{equation}

\section{Cosmological setting: Space-time with planar symmetry $\longrightarrow$
universe eccentricity $\longrightarrow$ polarization angle }

\subsection{Space-time anisotropy and magnetic energy density evolution}

Let us consider cosmological models with planar symmetry, i.e., having
a similar scale factor on the first two spatial coordinates. The most general
line-element of a geometry with plane-symmetry is \cite{taub}

 \begin{equation}\label{lineelement}
  ds^2=dt^2-b^2(dx^2+dy^2)-c^2 dz^2\,,
 \end{equation}

where $b(t)$ and $c(t)$ are the scale factors, which are normalized in order
that $b(t_0)=1=c(t_0)$ at the present time $t_0$. As Eq. (\ref{lineelement})
shows, the symmetry is on the (xy)-plane. The coherent temperature and
polarization patterns produced in homogeneous but anisotropic cosmological
models (Bianchi type with a Friedman-Robertson-Walker limit has been studied
in \cite{sung}).

The Christoffel symbols corresponding to the metric (\ref{lineelement}) are

\begin{equation}\label{Christoffel}
\Gamma^0_{11}=\Gamma^0_{22}=b{\dot b}\,, \quad \Gamma^0_{33}=c{\dot c}\,,
\end{equation}

 \[
 \Gamma^1_{01}=\Gamma^2_{02}=\frac{\dot b}{b}\,, \quad \Gamma^3_{03}=
\frac{\dot c}{c}\,.
 \]

The dot stands for derivative with respect to the cosmic time $t$.


To make an estimate on the parameter $\delta$, we have to investigate
in more detail the geometry with planar symmetry. As pointed out by
Campanelli-Cea-Tedesco (CCT) in \cite{cct}, the most general tensor
consistent with the geometry (\ref{lineelement}) is

 \[
 T^\mu_{\,\,\nu} = diag(\rho, -p_\parallel, -p_\parallel, -p_\perp)=T^\mu_{(I)
\nu}+T^\mu_{(A)\nu}\,,
 \]

in which $T^\mu_{(I)\nu}=diag(\rho, -p,-p,-p)$ is the standard isotropic
energy-momentum tensor describing matter, radiation, or cosmological constant,
and $T^\mu_{(A)\nu} = diag(\rho^A, -p^A, -p^A. -p^A)$ represents the anisotropic
contribution which induces the planar symmetry, and can be given by a uniform
magnetic field, a cosmic string, a domain wall \cite{cct-note}. In what follows,
we shall consider a Universe matter dominated ($p=0$) with planar symmetry
generated by a uniform magnetic field $B(t)$.

Magnetic fields have been observed in galaxies, galaxy clusters, and extragalactic
structures \cite{giovannini}, and it is assumed that they may have a primordial origin
\cite{primordialoriginB,kunze}. Due to the high conductivity of the primordial
plasma, the magnetic field evolves as $B(t)\sim b^{-2}$ being frozen into the
plasma \cite{parker,giovannini} (see below). Denoting with $\rho_B$
the magnetic field density, the energy-momentum tensor for a uniform magnetic
field can be written as $T^\mu_{(B) \; \nu} = \rho_B  diag(1, -1, -1, -1)$.

According to (\ref{energydensity1}), we find that the energy density of the
magnetic field is given by

\begin{equation}\label{energydensityB-delta}
 \rho_B = \frac{B^2}{8\pi}\left(\frac{B^2}{2\Lambda^4}\right)^{\delta-1}\,.
\end{equation}

The evolution law of the energy density $\rho_B$ is given by\cite{barrow-marteens-tsagas}

\begin{equation}\label{rhoBevolution}
{\dot \rho}_B+\frac{4}{3}\Theta \rho_B + 16\pi \sigma_{ab}\Pi^{ab} = 0 \,,
\end{equation}

where $\Theta$ is the volume expansion (contraction) scalar, $\sigma_{ab}$ is
the shear, and $\Pi^{ab}$ the anisotropic pressure of the fluid. In a highly
conducting medium we still have with good approximation $B\sim b^{-2}$ provided
that anisotropies can be neglected (this means that we neglect radiative effect
of the primordial fluid).


\subsection{Space-time eccentricity and polarization angle}

We shall assume that photons propagate along the (positive) $x$-direction, so
that $k^\mu=(k^0, k^1, 0, 0)$ \cite {Xinmin}. Gauge invariance assures that the polarization
vector of photons has only two independent components, which are orthogonal to
the direction of the photons motion.  Therefore, we are only interested in how the
components of the polarization vector  (ε2 and ε3) change.  It then follows that
$\Upsilon^\sigma$ defined in (\ref{Asigma}) assumes the form

 \begin{equation}\label{Asigmaexplicit}
  \Upsilon^\sigma=-k^0\left[\delta^{\sigma 2}\frac{\dot b}{b}\varepsilon^2
+\delta^{\sigma 3}\frac{\dot c}{c}\varepsilon^3 + \left(b{\dot b}(\varepsilon^2)^2
+c{\dot c}(\varepsilon^c)^2\right)\varepsilon^\sigma\right]
 \end{equation}

The components of $\Upsilon^\sigma$ given by (\ref{Asigmaexplicit}) vanish in
the case of a Friedman-Robertson-Walker geometry.

By defining the affine parameter $\lambda$ which measures the distance along
the line-element, $k^\mu\equiv dx^\mu/d\lambda$, one obtains that $\varepsilon^2$
and $\varepsilon^3$ satisfy the following geodesic equation (from Eq.
(\ref{k-nabla-epsilon}))


\begin{eqnarray}
& & \frac{d\varepsilon^2}{d\lambda} +\frac{\dot b}{b} k^0\varepsilon^2 =
-\frac{\delta-1}{\delta}k^0 \left[\frac{\dot b}{b}+b{\dot b}(\varepsilon^2)^2
+c{\dot c}(\varepsilon^3)^2 \right] \varepsilon^2 \nonumber \\
& & \frac{d\varepsilon^3}{d\lambda} + \frac{\dot c}{c}k^0 \varepsilon^3 =
-\frac{\delta-1}{\delta}k^0 \left[\frac{\dot c}{c}+ b{\dot b} (\varepsilon^2)^2
+c{\dot c}(\varepsilon^3)^2\right] \varepsilon^3 \nonumber
\end{eqnarray}

These equations can be further simplified if one observes that $k^0=dt/d\lambda$

\begin{eqnarray}
\frac{1}{k^0}{\cal D}\ln(b\varepsilon^2)=\frac{d\ln(b\varepsilon^2)}{dt} &=&
-\frac{\delta-1}{\delta}\left(-\frac{\dot
b}{b} +\frac{\dot c}{c}\right)(c\varepsilon^3)^2\,, \label{eqmotioneps2}\\
 \frac{1}{k^0}{\cal D}\ln(c\varepsilon^3) =  \frac{d\ln(c\varepsilon^3)}{dt} &=&
-\frac{\delta-1}{\delta}\left(-\frac{\dot
c}{c} +\frac{\dot b}{b}\right)(b\varepsilon^2)^2\,.
\label{eqmotioneps3}
\end{eqnarray}

where

 \begin{equation}\label{calD}
    {\cal D}\equiv k^\mu \nabla_\mu\,.
 \end{equation}

Moreover, the difference of the Hubble expansion rate ${\dot b}/b$ and ${\dot c}/c$
can be written as

 \begin{equation}\label{Hb-Hc}
    \frac{\dot b}{b}-\frac{\dot c}{c} = \frac{1}{2(1-e^2)}\frac{de^2}{dt}
 \end{equation}

where we have introduced the eccentricity

  \begin{equation}\label{eccentricity}
    e(t)=\sqrt{1-\left(\frac{c}{b}\right)^2}\,.
 \end{equation}

The polarization angle $\alpha$ is defined as $\alpha=\arctan[(c\varepsilon^3)/(b\varepsilon^2)]$.
 Its time evolution is governed by equation

\begin{equation}\label{Dalpha0}
{\cal D \alpha}- \frac{\delta-1}{2\delta}\,k^0  \left(\frac{\dot b}{b}-\frac{\dot c}{c}\right)
[(b\varepsilon^2)^3+(c\varepsilon^3)^3]=0\,.
 \end{equation}

However, Eqs. (\ref{eqmotioneps2}) and (\ref{eqmotioneps3}) implies that both $b\varepsilon^2$
and $c\varepsilon^3$ evolves as $A_{i}+ (\delta - 1) f_{i}(t)$, $i= 2, 3$ , where $f_i(t)$ is a
function of time and $A_i$ are constant of integration. Therefore, to leading order $(\delta-1)$
Eq. (\ref{Dalpha0})  reads

 \begin{equation}\label{Dalpha}
 {\cal D \alpha}-\frac{\delta-1}{\delta} \,\frac{K}{2} k^0  \left(\frac{\dot b}{b} - \frac{\dot c}
{c}\right) + {\cal O}((\delta-1)^2)=0\,.
 \end{equation}

where $K=A_2+A_3$.

To compute the rotation of the polarization angle, one needs to evaluate
$\alpha$ at two distinct instants. In the cosmological context that we are
considering is assumed that the reference time $t$ corresponds to the
moment in which photons are emitted from the last scattering surface,
and the instant $t_0$ corresponds to the present time. One, therefore,
gets \footnote{Preliminary calculations \cite{H-G_2010_C} performed
in terms of the electromagnetic field $F_{\mu\nu}$ and of time evolution
of the Stokes parameters $I, Q, U, V$ (this approach is alternative to one
presented in the Sec. II of the paper where the analysis is performed in
terms of the 4-potential $A_\mu$) yield again the result (\ref{rotationangle}).
Calculations show that the total flux $I$ is not the same along the three spatial
directions, as expected owing to the different expansion of the Universe along
the $x, y$ and $z$ directions. Moreover the time evolution of the Stokes parameters
turns out to be a mixture of each others, which reduce to standard results as $\delta=
1$. The polarization angle is defined as $2\alpha=\arctan(U/Q)$.}

 \begin{equation}\label{rotationangle}
 \Delta \alpha=\alpha(t)-\alpha(t_0)=\frac{\delta-1}{4\delta}Ke^2(z)\,,
 \end{equation}

where we have used $e(t_0)=0$ because of the normalization condition $b(t_0)
=c(t_0)=1$ and $\log (1-e^2)\sim -e^2$.

Notice that for $\delta=1$ or $e^2=0$ there is no rotation of the polarization angle,
as expected. Moreover, in the case in which photons propagate along the direction
$z$-direction, so that ${\bar k}^a = (\omega_0, 0, 0, k)$, we find that the NLED
 have no effects as concerns to the rotation of the polarization angle.

As arises from (\ref{rotationangle}), $\Delta \alpha$ vanishes in the
limit $\delta=1$, so that no rotation of the polarization angle occurs in
the standard electrodynamics, even if the background is described by a
geometry with planar symmetry. Moreover, even if $\delta \neq 1$, $\Delta
\alpha$ still vanishes for an isotropic and homogeneous cosmology described
by the Friedman-Robertson-Walker element line ($b=c$) $ds^2=dt^2-b^2(dx^2+
dy^2+dz^2)$, because in such a case the eccentricity vanishes (this agrees
with the fact that for this background the components of $\Upsilon^\sigma$,
Eq. (\ref{Asigmaexplicit}), are zero).

\subsection{Eccentricity evolution on cosmic time}

The time evolution of the eccentricity is determined from the Einstein field
equations

 \begin{equation}\label{eccentricityevol}
 \frac{1}{1-e^2}\frac{d(e{\dot e})}{dt}+3H_b (e{\dot e})+\frac{(e{\dot e})^2}
{(1-e^2)^2}=2\kappa \rho_B\,,
 \end{equation}

where $H_b={\dot b}/b$.

It is extremely difficult to exactly solve this equation. We shall therefore
assume that the $e^2$-terms can be neglected. Since $b(t)\sim t^{2/3}$ during
the matter-dominated era, Eq. (\ref{eccentricityevol}) implies

 \begin{equation}\label{eccentricitysol}
 e^2(z)=18 F_\delta (z) \Omega_B^{(0)}\,,
 \end{equation}

where we used $1+z=b(t_0)/b(t)$, $e(t_0)=0$, and

\begin{equation}\label{Fdelta}
    F_\delta \equiv \frac{3}{(9-8\delta)(4\delta-3)}-2-\frac{3(1+z)^{4
\delta-3}}{(9-8\delta)(4\delta-3)}+2(1+z)^{\frac{3}{2}}\,.
\end{equation}

$\Omega_B^{(0)}$ is the present energy density ratio

 \begin{equation}\label{OmegaB}
 \Omega_B^{(0)}=\frac{\rho_B}{\rho_{cr}}=\frac{B^2(t_0)}{8\pi
\rho_{cr}}\left(\frac{B^2(t_0)}{2\Lambda^4}\right)^{\delta-1}\simeq
10^{-11}\left(\frac{B(t_0)}{10^{-9}\rm G}\right)^2\left(\frac{B^2(t_0)}{2\Lambda^4}
\right)^{\delta-1}\, ,
 \end{equation}

with $\rho_{cr}=3H_b^2(t_0)/\kappa = 8.1 h^2 10^{-47}$ GeV$^4$ ($h=0.72$
is the little-$h$ constant), and $B(t_0)$ is the present magnetic field
amplitude.

From Eq. (\ref{rotationangle})  then follows

 \begin{equation}\label{rotationangle1}
 \Delta \alpha=\frac{\delta-1}{4\delta}\, K \, e^2 (z_{dec})\,.
 \end{equation}
where $e(z_{dec})^2$ the eccentricity (\ref{eccentricitysol}) evaluated at
the decoupling $z=1100$.

\subsection{Constraints on parameter $\Lambda$ from extragalactic ${\bf B}$
strengths in an ellipsoidal Universe}

To make an estimate on the parameter $\delta$, we need the order of amplitude
of the present magnetic field strength $B(t_0)$. In this respect, observations indicate that
there exist, in cluster of galaxies, magnetic fields with field strength $(10^{-7}-10^{-6})$
G on $10$ kpc - 1 Mpc scales, whereas in galaxies of all types and at cosmological distances,
the order of magnitude of the magnetic field strength is $\sim 10^{-6}$ G on (1-10) kpc scales.
The present accepted estimations is \cite{giovannini} \footnote{The bound (\ref{Bboundpresent})
is consistent with the estimation on the present value of the magnetic field strength obtained
from Big Bang Nucleosynthesis (BBN). As before pointed out, the magnetic fields scales as
$B\sim b^{-2}$ where the scale factor does depend on the temperature $T$ and on the total
number of effectively massless degree of freedom $g_{*S}$ as $b\propto g_{*S}^{-1/3}
T^{-1}$ \cite{kolbbook}. The upper bound on the magnetic field at the epoch of the  BBN is
given by \cite{grassoPLB} $B(T_{\rm{BBN}})\lesssim 10^{11}$G, where according to the
standard cosmology $T_{\rm{BBN}}=10^9$K$\simeq 0.1$MeV. Referred to the present
value of the magnetic field, the bound on $B(T_{\rm{BBN}})$ becomes \cite{rubinstein,cct}

 \begin{equation}\label{Bboundpresent}
    B(t_0)=\left(\frac{g_{*S}(T_0)}{g_{*S}(T_{\rm{BBN}})}\right)^{2/3}\left(\frac{T_0}
{T_{\rm{BBN}}}\right)^2   B(T_{\rm{BBN}})  \lesssim 6 \times 10^{-7}\rm{G}\,,
 \end{equation}

where $T_0=T(t_0)\simeq 2.35 \times 10^{-4}$eV and $g_{*S}(T_{\rm{BBN}})\simeq
g_{*S}(T_0)\simeq 3.91$ \cite{kolbbook}.}

 \begin{equation}\label{BvalueBound}
 B(t_0)\lesssim 10^{-9} \; \rm G\,.
 \end{equation}

\begin{figure}
{\includegraphics{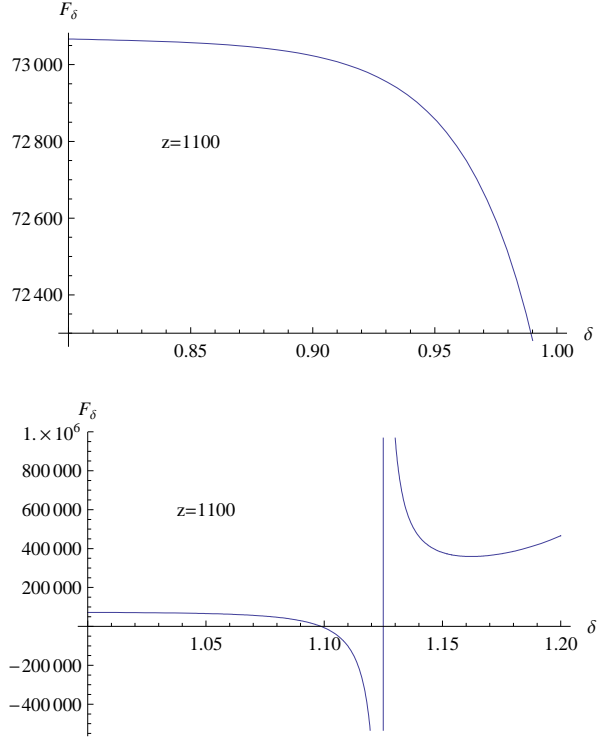}}
\caption{In this plot is represented $F_\delta$ vs $\delta$ for $\delta\leq 1$ (upper plot) and
$\delta\geq 1$ (lower plot). The condition that the eccentricity is positive follows for $F_\delta
>0$.}
 \label{Fig-z=1100}
\end{figure}

Moreover, for an ellipsoidal Universe the eccentricity satisfies the relation $0\leq e^2 < 1$. The
condition $e^2>0$ means $F_\delta>0$, with $F_\delta$ defined in (\ref{Fdelta}). The function
$F_\delta$ given by Eq. (\ref{Fdelta}) is represented in Fig. \ref{Fig-z=1100}. Clearly the
allowed region where $F_\delta$ is positive does depend on the redshift $z$. On the other hand,
the condition $e^2<1$ poses constraints on the magnetic field strength. By requiring $e^2<
10^{-1}$ (in order that our approximation to neglect $e^2$-terms in (\ref{eccentricityevol})
holds), from Eqs. (\ref{eccentricitysol})-(\ref{OmegaB}) it follows

 \begin{equation}\label{Bvalueecc}
 B(t_0)\lesssim 9 \times 10^{-8} \rm G\,.
 \end{equation}


It must also be noted that such magnetic fields does not affect the expansion rate of the universe
and the CMB fluctuations because the corresponding energy density is negligible with respect to
the energy density of CMB.

\section{Light propagation in NLED and birefringence}

In this Section we discuss the modification of the light velocity (birefringence effect) for the
model of nonlinear electrodynamics $L(X, Y)$. We shall follow the paper \cite{obukhov} (see
also \cite{plebanski,metricoptic}), in which is studied the propagation of wave in local nonlinear
electrodynamics by making use of the Fresnel equation for the wave covectors $k_\mu$. The
latter are related to phase velocity $v$ of the wave propagation by the relation $k_i=\displaystyle
{\frac{k_0}{v}{\hat k}_i}$, where ${\hat k}_i$ are the components of the unit 3-covector. Thus,
in what follows we confine ourselves to the phase velocity.  It is straightforward to show that for
the models under consideration (\ref{L(X)}) the group velocity is always greater or equal to the
phase velocity \cite{obukhov}.

The main result in Ref. \cite{obukhov} corresponds to the {\it optic metric} tensors

\begin{equation}\label{g1}
 g_1^{\mu\nu}={\cal X} g^{\mu\nu}+({\cal Y}+\sqrt{{\cal Y}-{\cal X}{\cal Z}})t^{\mu\nu}\,,
 \end{equation}

 \begin{equation}\label{g2}
 g_2^{\mu\nu}={\cal X} g^{\mu\nu}+({\cal Y}-\sqrt{{\cal Y}-{\cal X}{\cal Z}})t^{\mu\nu}\,,
 \end{equation}

which describe the effect of birefringent light propagation in a generic model for nonlinear
electrodynamics. The quantities ${\cal X}$, ${\cal Y}$, and ${\cal Z}$ are related to the
derivatives of the Lagrangian $L(X, Y)$ with respect to the invariant $X$ and $Y$, and $t^{
\mu\nu}=F^{\mu\alpha} F^\nu_{\,\,\alpha}$.

For our model, expressed byEq. (\ref{L(X)}),  the quantities ${\cal X}$, ${\cal Y}$, and ${
\cal Z}$ are given by

 \[
  {\cal X}\equiv K_1^2= \frac{\gamma^2 \delta^2}{4}  X^{2(\delta-1)}\,, \quad {\cal Y}\equiv
K_1 K_2=\frac{\gamma^2\delta^2}{4}\, (\delta-1)X^{2(\delta-1)-1}\,, \quad {\cal Z}=0\,,
  \]

where $K_1= \displaystyle{4\frac{\partial L}{\partial X}}$ and $K_2= 8\displaystyle{\frac{
\partial^2 L}{\partial X^2}}$, while the metrics (\ref{g1}) and (\ref{g2})
are

  \[
  g^{\mu\nu}_1=K_1(K_1 g^{\mu\nu}+2K_2 t^{\mu\nu})\,, \quad g^{\mu\nu}_2=K_1^2
g^{\mu\nu}.
  \]

As a consequence, birefringence is present in our model. This means that some photons
propagate along the standard null rays of spacetime metric $g^{\mu\nu}$, whereas other
photons propagate along rays null with respect to the optical metric $K_1 g^{\mu\nu} +2
K_2 t^{\mu\nu}$.

The velocities of the light wave can be derived by using the light cone equations (effective
metric)

\[
 g_1^{\mu\nu}k_\mu k_\nu = 0 \quad {\rm and} \quad  g_2^{\mu\nu}k_\mu k_\nu = 0\,.
 \]

It is worthwhile to report the general expression for the average  value of the velocity scalar
\cite{obukhov}

 \[
 \langle v^2 \rangle= 1+\frac{4}{3}  \frac{T^{00}({\cal Y}+{\cal Z} t^{00})}{{\cal X}+2{\cal
Y}t^{00}+{\cal Z}(t^{00})^2}+  \frac{2}{3}{\bf S}^2\, \frac{2{\cal Y}^2-{\cal X}{\cal Z} +
{\cal Z}(t^{00})^2+2{\cal Y}{\cal Z}t^{00}}{[{\cal X}+2{\cal Y}t^{00}+{\cal Z}(t^{00})^2]^2}
 \]

where $T^{00} = - t^{00}+X = ({\bf E}_\gamma^2 + {\bf B}_\gamma^2)/2$ ($t^{00} = -{\bf
E}_\gamma^2$), and ${\bf S}^2=\delta_{\mu\nu} t^{0\mu}t^{0\nu}$, where ${\bf S}={\bf E}
\times {\bf B}$ is the energy flux density. The subscript $\gamma$ is introduced for distinguishing
the photon field from the magnetic background. The value of the mean velocity has been derived
averaging over the directions of propagation and polarization. For our model, we get

 \begin{equation}\label{vourmodel}
   \langle v^2 \rangle \simeq 1+(\delta-1) R + (\delta-1)^2 S\,,
 \end{equation}

  \[
   R\equiv \frac{4}{3}\frac{T^{00}}{4X+2(\delta-1)t^{00}} \,, \quad S=\frac{4}{3}
\frac{{\bf S}^2}{[4X+2(\delta-1)t^{00}]^2}
  \]

The high accuracy of optical experiments in laboratories requires tiny deviations from standard
electrodynamics. This condition is satisfied provided $|\delta-1|\ll 1$. Moreover, there are
two aspects related to (\ref{vourmodel}):

 \begin{itemize}
 \item The average velocity does depend on ({\it only}) the parameter $\delta$, so that $\gamma$
or $\Lambda$ in our model can be fixed independently. This task is addressed in the next Section.
 \item Because $R$ is positive, one has to demand that $\delta-1 <0$ in order that $v^2<1$.
 \end{itemize}

The above considerations hold for flat spacetime, and can be straightforwardly generalized to the
case of curved space time \cite{obukhov}.

\section{Stokes parameters, rotated CMB spectra and constraints on parameter
$\Lambda$}

The propagation of photons can be described in terms of the Stokes parameters
$I$, $Q$, $U$, and $V$. The parameters $Q$ and $V$ can be decomposed in
gradient-like ($G$) and a curl-like ($C$) components \cite{stebbins} ($G$
and $C$ are also indicated in literature as $E$ and $B$), and characterize
the orthogonal modes of the linear polarization (they depend on the axes
where the linear polarization are defined, contrarily to the physical
observable $I$ and $V$ which are independent on the choice of coordinate
system).

The polarization $G$ and $C$ and the temperature ($T$) are crucial because
they allow to completely characterize the CMB on the sky.  If the Universe is
isotropic and homogeneous and the electrodynamics is the standard one, then
the $TC$ and $GC$ cross-correlations power spectrum vanish owing to the
absence of the cosmological birefringence. In presence of the latter,
on the contrary, the polarization vector of each photons turns out to be
rotated by the angle $\Delta \alpha$, giving rise to $TC$ and $GC$ correlations.

Using the expression for the power spectra $C_l^{XY}\sim \int dk [k^2 \Delta_X(
t_0) \Delta _Y(t_0)]$, where $X, Y=T, G, C$ and $\Delta_{X}$ are the polarization
perturbations whose time evolution is controlled by the Boltzman equation, one
can derive the correlation for $T$, $G$ and $C$ in terms of $\Delta \alpha$
\cite{amelinojcap}\footnote{Notice that in Ref.\cite{lue} the analysis did not
include the rotation of the CMB spectra, and in Ref.\cite{stebbins} the analysis
focused on only the TC and TG modes. Other approximated approaches to discuss the
rotation angle can be found in Refs.\cite{hu,feng}.}

\begin{equation}
 C_l^{\prime\,TC}= C_l^{TC}\sin 2\Delta \alpha\,, \quad   C_l^{\prime\,TG}=
C_l^{TG}\cos 2\Delta \alpha\,,
\end{equation}

 \begin{equation}
 C_l^{\prime\,GC}= \frac{1}{2}\left(C_l^{GG}-C_l^{CC}\right)\sin 4\Delta \alpha\,,
\end{equation}

\begin{equation}
 C_l^{\prime\,GG}= C_l^{GG}\cos^2 2\Delta \alpha+C_l^{CC}\sin^2 2\Delta \alpha\,,
\end{equation}

\begin{equation}
C_l^{\prime\,CC}= C_l^{CC}\cos^2 2\Delta \alpha+C_l^{GG}\sin^2 2\Delta \alpha\,.
\end{equation}

The prime indicates the rotated quantities. Notice that the CMB temperature
power spectrum remains unchanged under the rotation.

Experimental constraints on $\Delta \alpha$ have been put from the observation
of CMB polarization by WMAP and BOOMERanG \cite{amelinojcap,xia}.

\begin{equation}\label{Deltaalphaconstraint}
 \Delta \alpha=(-2.4 \pm 1.9)^\circ = [-0.0027\pi, -0.0238\pi]\,.
\end{equation}

The combination of Eqs. (\ref{Deltaalphaconstraint}) and (\ref{rotationangle1}),
and the laboratory constraints $|\delta-1| \ll 1$ allow to estimate $\Lambda$.

\subsection{Estimative of $\Lambda$}

To estimate $\Lambda$ we shall write

 \begin{eqnarray}\label{Bb}
    B &=& 10^{-9+b}\, {\rm G} \qquad \,\, \, b\lesssim 2\,, \\
    F_\delta &=& 2 z^{3/2} \qquad \qquad z =1100 \gg 1\,.
 \end{eqnarray}

The bound (\ref{Deltaalphaconstraint}) can be therefore rewritten in the form

 \begin{equation}\label{boubddelta}
    \frac{10^{-3}}{\cal A}\lesssim |\delta-1|\lesssim \frac{10^{-2}}{\cal A}\,,
 \end{equation}

where

 \begin{equation}\label{calA}
 {\cal A} \equiv \frac{9K}{14}\, F_\delta \Omega_B^{(0)}\simeq K\, 10^{-6+2b}\left[0.24
\times 10^{-56+2b}  \left(\frac{\rm GeV}{\Lambda}\right)^4\right]^{\delta-1}\,.
 \end{equation}

The condition $|\delta-1|\ll 1$ requires ${\cal A}\gg 1$. It turns out convenient to set

 \begin{equation}\label{calA1}
    {\cal A}=10^a \,, \qquad a > {\cal O}(1)\,.
 \end{equation}

From Eqs. (\ref{calA}) and (\ref{calA1}) it then follows

 \begin{equation}\label{Lambda}
    \Lambda = 10^{-14 + b/2} \left[\frac{1}{K}\, 10^{a-2b+6}\right]^{-\frac{1}{4(\delta-1)}}
\rm GeV\,,
 \end{equation}

or equivalently

 \begin{equation}\label{LogLambda}
    {\rm Log} \left[\frac{\Lambda}{\rm GeV}\right] = \left(-14 + \frac{b}{2}\right)+\frac{(-1)}
{4(\delta-1)} \left[a-2b+6-{\rm Log}K\right]\,.
 \end{equation}

The constant $K$ can now be determined to fix the characteristic scale $\Lambda$. Writing
$\Lambda = 10^{\Lambda_x}$ GeV, where $\Lambda_{x=Pl} = 19$, $\Lambda_{\rm GUT} =
16$ and $\Lambda_{\rm EW} = 3$ for the Planck, GUT and electroweak (EW) scales,
respectively,  Eq. (\ref{LogLambda}) yields

\begin{equation}\label{K}
K=10^{a-2b+6-\zeta}\,, \qquad \zeta\equiv \frac{4(\delta-1) \Lambda_x}{14-b/2}\ll 1\,.
\end{equation}

\begin{figure}
\resizebox{15cm}{!}{\includegraphics{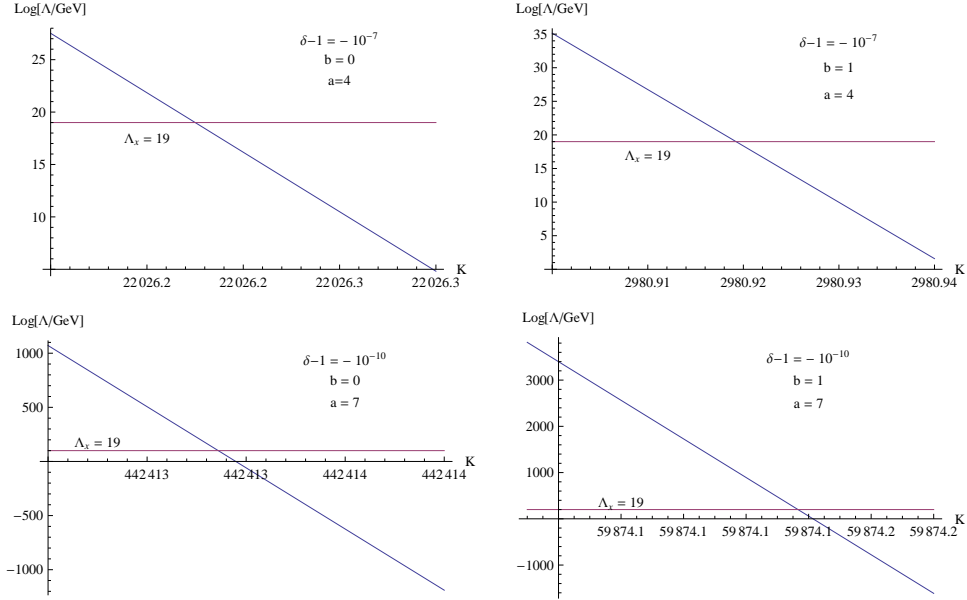}}
\caption{$\Lambda$ vs $K$ for different values of the parameter $\delta-1$, $a$ and $b$. The
parameter $a$ is related to the range in which $\delta-1$ varies, i.e. $-10^{-3-a}\lesssim \delta-1
\lesssim -10^{-2-a}$, while $b$ parameterizes the magnetic field strength $B=10^{-9+b}$ G.
The red-shift is $z=1100$. Plot refers to Planck scale $\Lambda=10^{\Lambda_x}$ GeV, with
$\Lambda_{x=Pl}=19$. Similar plots can be also obtained for GUT ($\Lambda_{\rm GUT} =
16$) and EW ($\Lambda_{\rm EW}=3$) scales.}
 \label{Fig2}
\end{figure}

In Fig. \ref{Fig2} is plotted Log$(\Lambda$/GeV) vs $K$ for fixed values of the parameters
$a$,  $b$ and $\delta-1$. Similar plots can be derived for GUT and EW scales

\section{Discussion and closing remarks}

In conclusion, in this paper we have calculated, in the framework of the nonlinear
electrodynamics, the rotation of the polarization angle of photons
propagating in a Universe with planar symmetry. We have found that
the rotation of the polarization angle does depend on the parameter
$\delta$, which characterizes the degree of nonlinearity of the
electrodynamics. This parameter can be constrained by making use
of recent data from WMAP and BOOMERang. Results show that the CMB
polarization signature, if detected by future CMB observations,
would be an important test in favor of models going beyond the standard
model, including the nonlinear electrodynamics.

Some comments are in order. In our investigation we have assumed that the
planar-symmetry is induced by a magnetic field. This is not the unique
case. In fact, a planar geometry can also be induced by topological
defects, such as cosmic string ($cs$) or domain wall ($dw$) \cite{cct}.
In such a case, one has \cite{cct}
 \begin{equation}
 e^2\Biggl|_{dw}=\frac{2}{7}\Omega_{dw}^{(0)}\left[\frac{3}{(1+z)^2}+
4(1+z)^{3/2} -7\right]\,,
  \end{equation}
and
  \begin{equation}
  e^2\Biggl|_{cs}=\frac{4}{5}\Omega_{cs}^{(0)}\left[\frac{3}{(1+z)}+2(1+z)^{3/2}
-5\right]\,,
  \end{equation}
where $\Omega_{(dw,cs)}^{(0)}$ are the present energy densities, in units
of critical density, of the domain wall and cosmic string. At the decoupling,
one obtains
  \begin{equation}
  e^2(z_{dec})\Biggl|_{dw}\simeq 10^{-4}\frac{\Omega_{dw}^{(0)}}{5\times 10^{-7}}\,,
   \end{equation}
and
  \begin{equation}
  e^2(z_{dec})\Biggl|_{cs}\simeq 10^{-4}\frac{\Omega_{cs}^{(0)}}{4\times 10^{-7}}\,.
  \end{equation}
The analysis leading to determine the bounds on $\delta$ from CMB polarization goes
along the line above traced.

Moreover, a complete analysis of the planar-geometry
is required to fix the parameter $\delta$. From a side, in our calculations in fact we have
assumed that the Universe is matter dominated. A more precise calculation should
require to use (to solve (\ref{eccentricityevol})) the relation
 \begin{equation}
t= \displaystyle{\frac{1}{H_0} \int_0^z \frac{1+z}{\sqrt{\Omega_m(1+z)^3+
\Omega_\Lambda}}\, dz}\; ,
 \end{equation}
where $\Omega_m =0.3$, $\Omega_\Lambda=0.7$ and $H_0=72$ km sec$^{-1}$
Mpc$^{-1}$ ($z=1100$). From the other side, a complete study of the Eq. (\ref{eccentricityevol})
is necessary in order to put stringent constraints on the parameter $\delta$.

As closing remark, we would like to point out that the approach to
analyze the CMB polarization in the context of NLED that we have
presented above can also be applied to discuss the extreme-scale
alignments of quasar polarization vectors \cite{H-L-et-al_2005}, a cosmic
phenomenon that was discovered by Hutsemekers\cite{H_1998} in the
late 1990's, who presented paramount evidence for very large-scale
coherent orientations of quasar polarization vectors (see also
Hutsemekers and Lamy \cite{H-L_2000}).\footnote{Hutsemekers and
Lamy, and collaborators, have presented, in a long series of papers
(not all cited here) published over the period 1998 to 2008, a
tantamount evidence
that the alignment of quasar polarization vectors is a factual
cosmological enigma deserving to be properly addressed in the
framework of the standard model of cosmology. The papers quoted
here are intended to call to the attention of attentive readers
the paramount evidence presenting this cosmic phenomenon.}. As
far as the authors
of the present paper are awared of, the issue has remained as an open
cosmological connundrum, with a few workers in the field having focused
their attention on to those intriguing observations. Nonetheless, we
quote ``en passant'' that in a recent paper \cite{H_et+al+2008}
Hutsemekers et al. discussed the possibility of such phenomenon
to be understood by invoking very light pseudoscalar particles
mixing with photons. They claimed that the observations of a sample
of 355 quasars with significant optical polarization present strong
evidence that quasar polarization vectors are not randomly oriented
over the sky, as naturally expected. Those authors suggest that the
phenomenon can be understood in terms of a cosmological-size effect,
where the dichroism and birefringence predicted by a mixing between
photons and very light pseudoscalar particles within a background
magnetic field can qualitatively reproduce the observations. They
also point out at a finding indicating that circular polarization
measurements could help constrain their mechanism.

Since cosmic magnetic fields have a typical strength of $\sim
10^{-7}-10^{-8}$ G, on average, for a characteristic distance
scale of 10-30 Mpc, it is our view that such phenomenology can be
understood in the framework of a nonlinear description of photon
propagation (NLED) over cosmic background magnetic fields and the
use of a planar symmetry for the space-time. Specifically, phenomena
involving light propagation as dichroism and birefringence can be
inscribed on to the framework of Heisenberg-Euler NLED, which
predicts the occurrence of birefringence on cosmological distance
scales. We plan to present such analysis in a forthcoming communication
\cite{H-G_2010_C}.

\vspace{0.2in}

{\bf Acknowledgments}: The authors express their gratitude to the referee
for relevant comments. The authors also thank Prof. Xin-min Zhang,
and Dr. Mingzhe Li for reading the manuscript and their suggestions.
H.J.M.C. thanks ICRANet International Coordinating Center, Pescara,
Italy for hospitality during the early stages of this work. G.L. acknowledges
the financial support of MIUR through PRIN 2006 Prot. $1006023491_{-}
003$, and of research funds provided by the University of Salerno.


\begin{thebibliography}{10}
\bibitem{cct-note} Notice that the cases discussed in \cite{cct}, i.e. the
magnetic fields or the topological defects as responsible of the planar
symmetry of the Universe, allow to explain the anomaly concerning the low
quadruple amplitude probed by WMAP. The interesting analysis performed by
CCT leads to the conclusion that an eccentricity at the decoupling of the
order of $e\sim 10^{-2}$ is indeed able to explain the drastic reduction
in the quadruple anisotropy without affecting higher multiples of the angular
power spectrum of the temperature anisotropies.
\end{thebibliography}
\end{document}